\documentstyle[prl,graphicx,aps]{revtex}
\begin{document}
	
\twocolumn[\hsize\textwidth\columnwidth\hsize\csname@twocolumnfalse\endcsname

\title{Higgs Driven Geodetic Evolution/Nucleation of de-Sitter Brane}
\author{Aharon Davidson, David Karasik and Yoav Lederer}
\address{Physics Department, Ben-Gurion University of the Negev,
Beer-Sheva 84105, Israel\\
\textbf{davidson@bgumail.bgu.ac.il}}
\maketitle
\begin{abstract}
    Geodetic evolution of a de-Sitter brane is exclusively
    driven by a Higgs potential, rather than by a plain
    cosmological constant.
    The deviation from Einstein gravity, parameterized  by
    the conserved bulk energy, is characterized by a hairy
    horizon which serves as the locus of unbroken symmetry.
    The quartic structure of the potential, singled out on
    finiteness grounds of the total (including the dark
    component) energy density, chooses the no-boundary proposal.
\end{abstract}
\pacs{PACS numbers: }
]    

The Randall-Sundrum model\cite{RS} has re-ignited the interest
in brane gravity.
A somewhat different approach, to be referred to as geodetic
brane gravity, has been advocated long ago by
Regge-Teitelboim\cite{RT}.
The corresponding geodetic field equations
\begin{equation}
    E^{\mu\nu}\left(y^{A}_{;\mu\nu}+
    \Gamma^{\,A}_{BC}y^{B}_{,\mu}y^{C}_{,\nu}\right)=0
\end{equation}
describe a generalized geodetic motion of an embedded lower
dimensional brane\cite{E,GW}, parameterized by means of
$x^{\mu}\,(\mu=0,1,\ldots,n)$, in a higher dimensional
background spanned by $y^{A}\,(A=0,1,\ldots,N)$.
The Einstein tensor
\begin{equation}
    E^{\mu\nu}\equiv\frac{1}{8\pi G}
    \left(R^{\mu\nu}-\frac{1}{2}g^{\mu\nu}R\right)-T^{\mu\nu}
\end{equation}
keeps track of the underlying standard Einstein-Hilbert
Lagrangian on the brane.
Non-conventionally, however, in the spirit of string/particle
theory, it is now the embedding vector $y^{A}(x^{\mu})$, rather
than the induced metric tensor
$g_{\mu\nu}=\eta_{AB}(y)y^{A}_{,\mu}y^{B}_{,\nu}$, which is
elevated to the level of the canonical gravitational field.
No wonder every solution of Einstein equations, that is
$E^{\mu\nu}=0$, is necessarily a solution of the corresponding
Regge-Teitelboim equations.
Furthermore, owing to the powerful identity
$\eta_{AB}y^{A}_{,\mu}\left(y^{B}_{;\nu\lambda}+
\Gamma^{\,B}_{CD}y^{C}_{,\nu}y^{D}_{,\lambda}\right)=0$,
one still automatically recovers energy-momentum conservation
$T^{\mu\nu}_{\,;\nu}=0$.
The case of a flat Minkowski background is favored on
various theoretical grounds\cite{RTgr}. 

\medskip
Within the framework of geodetic brane cosmology,
formulated by virtue of $5$-dimensional local isometric
embedding, only a single independent RT-equation survives,
namely $\frac{d}{dt}\left(\sqrt{-g}E^{tt}\dot{y}^{0}\right)=0$.
A trivial integration gives then rise to
\begin{equation}
    \rho a^{3}(\dot{a}^{2}+k)^{1/2}-
    3a(\dot{a}^{2}+k)^{3/2}=-\frac{\omega}{\sqrt{3}} ~,
\end{equation}
accompanied by $\dot{\rho}+3\frac{\dot{a}}{a}(\rho+P)=0$.
The constant of integration $\omega$, recognized as
the conserved bulk energy conjugate to the cyclic embedding time
coordinate $y^{0}(t)$, parameterizes the deviation from the Einstein
limit (where $\dot{a}^{2} + k\rightarrow\frac{1}{3}\rho a^{2}$).
A physicist equipped with the traditional Einstein formalism,
presumably unaware of the underlying RT physics, would naturally
re-organize the latter equation into
\begin{equation}
    \dot{a}^{2}+k =
    \frac{1}{3}\left(\rho+\Delta\rho\right)a^{2} ~,
\end{equation}
squeezing all 'anomalous' pieces into $\Delta\rho$.
Our physicist may rightly conclude\cite{AD} that the FRW
evolution of the Universe is governed by the effective
$\rho+\Delta\rho$ rather than by the primitive $\rho$, and
thus may further identify (or just use the language of)
$\Delta\rho \equiv\rho_{dark}$.
A simple algebra reveals the cubic consistency relation
\begin{equation}
    \left(\rho+\rho_{dark}\right)\rho_{dark}^{2}=
    \frac{\omega^{2}}{a^{8}} ~.
    \label{rhod}
\end{equation}
Here, we focus attention on the prototype case involving a
minimally coupled scalar field $\phi(t)$, subject to
\begin{equation}
    \ddot{\phi}+3\frac{\dot{a}}{a}\dot{\phi}+
    \frac{dW(\phi)}{d\phi} = 0 ~.
\end{equation}

\medskip
Two exclusive features of geodetic brane cosmology, relevant
for our discussion, are worth noting, namely

\noindent (i) \textit{Positive definite total energy density:}
To be specific, eq.(\ref{rhod}) tells us that $\rho_{total}
\equiv \rho+\rho_{dark} \geq 0$.
Curiously, this conclusion holds even for $\rho<0$.
    
\noindent (ii) \textit{Cosmic duality:}
FRW evolution cannot tell the configuration $\{\rho,\,\rho_{dark}\}$
from its dual $\{\rho+2\rho_{dark},\,-\rho_{dark}\}$, both sharing a
common $\rho_{total}$.
A pedagogical example can be provided by the 'empty' $\rho=0$ case,
whose dual happens to constitute a scalar field theory governed by a
quintessence-type potential.

\medskip
To uncover the mysteries of the dark component $\rho_{dark}$, we
start by asking a simple minded question that has a well established
answer within Einstein cosmology.
Namely, \textit{under what conditions can one obtain eternal deSitter
evolution?}
It is well known that Einstein gravity requires the introduction of a
positive cosmological constant $\rho=\Lambda>0$.
Here, however, adopting (say) the $k>0$ case, we are after the
tenable scalar potential $W(\phi)$, if any, capable of supporting
$\rho_{total}=\Lambda>0$.

\medskip
Guided by eq.(\ref{rhod}), the crucial thing to notice
is the split
\begin{equation}
    \rho = \Lambda + \frac{\omega}{\Lambda^{1/2}a^{4}} ~, \quad
    \rho_{dark} = -\frac{\omega}{\Lambda^{1/2}a^{4}} ~.
\end{equation}
Differentiating $\rho= \frac{1}{2}\dot{\phi}^{2} + W(\phi)$, we substitute
$\frac{dW}{d\phi}+\ddot{\phi}$ by $-3\frac{\dot{a}}{a}\dot{\phi}$
to learn that
\begin{equation}
    \dot{\phi}^{2} = \frac{4\omega}{3\Lambda^{1/2}a^{4}} ~,
\end{equation}
and appreciate the fact that associated with our $\omega>0$ is a
\textit{negative} dark energy component $\rho_{dark}<0$ (recall
in passing the existence of a yet unspecified $\rho_{dark}>0$ dual
theory with identical brane evolution).
We also find that
\begin{equation}
    W = \Lambda +\frac{\omega}{3\Lambda^{1/2}a^{4}} ~,
\end{equation}
and would like, in search of a differential equation for
$W(\phi)$, to also express $\frac{dW}{d\phi}$ as a parametric
function of $a$.
To do so, we calculate $\ddot{\phi}$ and plug the result into
the scalar field equation to obtain
\begin{equation}
    \frac{dW}{d\phi} = \pm \frac{1}{a^{3}}
    \sqrt{\frac{4\omega}{3\Lambda^{1/2}}
    (\frac{1}{3}\Lambda a^{2}-k)} ~.
\end{equation}
Defining now $f\equiv \frac{1}{3}\Lambda-
\sqrt{\frac{3k^{2}\Lambda^{1/2}}{\omega}(W-\Lambda)}$,
it satisfies
\begin{equation}
    \left(\frac{df}{d\phi}\right)^{2} =
    \frac{3k^{2}\Lambda^{1/2}}{\omega}f ~.
\end{equation}
The solution of this equation is rather serendipitous:
\textit{The unique scalar potential capable of supporting
an inflationary deSitter brane is a Higgs potential},
given explicitly by
\begin{equation}
    \fbox{${\displaystyle W(\phi) =
    \Lambda + \frac{3\Lambda^{1/2}k^{2}}{16\omega}
    \left(\phi^{2} -\frac{4\omega\Lambda^{1/2}}{9k^{2}}
    \right)^{2}}$}
    \label{Higgs}
\end{equation}
Associated with this potential, but relying on certain
creation initial conditions (to be specified soon) is the
full $k>0$ classical solution
\begin{mathletters}
    \begin{eqnarray}
	a(t) & = & \sqrt{\frac{3k}{\Lambda}}
	\cosh{\sqrt{\frac{\Lambda}{3}}t} ~, \\
	\phi(t) & = & \sqrt{\frac{4\omega\Lambda^{1/2}}{9k^{2}}}
	\tanh{\sqrt{\frac{\Lambda}{3}}t} ~.
    \end{eqnarray}
    \label{solution}
\end{mathletters}
On symmetry (and forth coming Euclidean) grounds, we find
it rewarding to follow Hartle and Hawking\cite{HH} and define
the proper scalar field $b(t) \sim a(t)\phi(t)$, describing
evolution by the hyperbola
\begin{equation}
    a(t)^{2}-b(t)^{2}=\frac{3k}{\Lambda} ~.
    \label{hyperbola}
\end{equation}

The emerging deSitter inflationary scheme, accompanied by
the auxiliary scalar field, deviates conceptually from the
conventional prescription.
Created with a radius of $a_{0}=\sqrt{\frac{3k}{\Lambda}}$
while sitting at the top of the hill $W_{0} = \Lambda\left(
1+\frac{\omega\Lambda^{1/2}}{27k^{2}}\right)$,
the exponentially growing brane slides down the potential
towards the absolute minimum conveniently located at the
Einstein limit $W_{\infty}=\Lambda$.
The scalar field, at the meantime, recovering from the
non-conventional creation initial conditions
\begin{equation}
    \phi_{0} = 0 ~,\quad
    \dot{\phi}_{0} = \sqrt{\frac{4\omega
    \Lambda^{3/2}}{9k^{2}}} ~,
\end{equation}
grows monotonically on its way to eventually picking up the
vacuum expectation value
\begin{equation}
    \langle\phi\rangle \equiv V =
    \sqrt{\frac{4\omega\Lambda^{1/2}}{9k^{2}}} ~.
\end{equation}
Altogether, accompanied by a seesaw-type $\rho\leftrightarrow
\rho_{dark}$ interplay, deSitter inflation is described within the
framework of geodetic brane cosmology as a spontaneously
symmetry breaking process, with Einstein gravity recovered at
the absolute minimum.
On the practical side, there is no need to artificially engineer the
shape of a slow-rolling scalar potential in order to maximize the
inflation period; an ordinary Higgs potential can do.

\medskip
Two important remarks are in order:

\noindent (i) For $k<0$, the situation is very much alike.
Truly, this time one faces
$a(t)\sim\sinh{\sqrt{\frac{\Lambda}{3}}t}$ and
$\phi(t) \sim \coth{\sqrt{\frac{\Lambda}{3}}t}$, but
the Higgs potential stays invariant under $k\rightarrow-k$.
Nucleated with size zero, accompanied by a monotonically
decreasing scalar field, our exponentially growing open
brane slides again towards the $W_{\infty}=\Lambda$
Einstein limit.
However, contrary to the closed $k>0$ case where only the
\textit{inner} section $(0\leq\phi\leq V)$ of the potential was
involved, it is the \textit{outer} section $(V\leq\phi<\infty)$
which participates in the $k<0$ game.
For $k=0$, the situation is less complicated, with the Higgs
potential reducing to a simple mass term.

\noindent (ii) The deSitter metric can also take the static
radially symmetric form
\begin{equation}
    ds^{2} = -\left(1-\textstyle{\frac{1}{3}}
    \Lambda R^{2}\right)dT^{2} +
    \frac{dR^{2}}{\left(1-\frac{1}{3}\Lambda R^{2}\right)} +
    R^{2}d\Omega^{2} ~,
\end{equation}
exhibiting an event horizon at $R=\sqrt{\frac{3}{\Lambda}}$.
Reflecting the seesaw interplay between the primitive and the
dark energy densities, the auxiliary scalar field plays here an
apparently paradoxical \textit{non-static} role.
To see the point, consider (say) the patch
$R\leq\sqrt{\frac{3}{\Lambda}}$ covered by
\begin{mathletters}
\begin{eqnarray}
    & \displaystyle{\sqrt{\frac{\Lambda}{3k}}R =
        r \cosh\sqrt{\frac{\Lambda}{3}}t ~,}& \\
    & \displaystyle{\coth \sqrt{\frac{\Lambda}{3}}T =
    \sqrt{1-kr^{2}} \coth\sqrt{\frac{\Lambda}{3}}t ~.}&
\end{eqnarray}
\end{mathletters}
In this coordinate system, the auxiliary $T$-dependent scalar
field acquires the form
\begin{equation}
    \fbox{$\displaystyle{\phi(T,R) = \frac
    {V\sqrt{1-\frac{1}{3}\Lambda R^{2}}
    \sinh\sqrt{\frac{\Lambda}{3}}T}
    {\sqrt{1+\left(1-\frac{1}{3}\Lambda R^{2}\right)
    \sinh^{2}\sqrt{\frac{\Lambda}{3}}T}}}$}
\end{equation}
giving rise to double-kink configuration (a kink-antikink
configuration for $R\geq\sqrt{\frac{3}{\Lambda}}$) \textit{scalar hair}.
In almost every point $R$ in space, elegantly avoiding the no-hair 
theorems of general relativity, the scalar field connects
$\phi(-\infty,R)\rightarrow -V$ with $\phi(\infty,R)\rightarrow V$.
It is exclusively on the event horizon, however, where the scalar
field, experiencing an infinite gravitational red-shift, gets
frozen in its unbroken phase!
In other words, \textit{the hairy event horizon appears as the
locus of unbroken symmetry}.

\medskip
To enter the Euclidean regime we perform the Wick rotation
$t \rightarrow 
-i\left(\tau-\frac{\pi}{2}\sqrt{\frac{3}{\Lambda}}\right)$.
The exact $k>0$ solution eq.(\ref{solution}) transforms then
into
\begin{mathletters}
    \begin{eqnarray}
	a(t) & \rightarrow & a_{E}(\tau) =
	\sqrt{\frac{3k}{\Lambda}}
	\sin{\sqrt{\frac{\Lambda}{3}}\tau} ~, \\
	\phi(t) & \rightarrow & i\phi_{E}(\tau) =
	i\sqrt{\frac{4\omega\Lambda^{1/2}}{9k^{2}}}
	\cot{\sqrt{\frac{\Lambda}{3}}\tau} ~.
    \end{eqnarray}
    \label{Esolution}
\end{mathletters}
The fact that the scalar field turns purely imaginary puts
us in a less familiar territory, highly reminding us of the
Coleman-Lee\cite{CL} scheme.
The imaginary time evolution is then best described by the
circle
\begin{equation}
    a_{E}^{2}(\tau) + b_{E}^{2}(\tau) = \frac{3k}{\Lambda} ~,
    \label{circle}
\end{equation}
recognized as the analytic continuation of eq.(\ref{hyperbola}).
This makes the familiar deSitter Euclidean time periodicity
$\Delta \tau = 2\pi\sqrt{\frac{3}{\Lambda}}$ manifest, and
opens the door for a generalized Hawking-Hartle no-boundary
proposal.

\begin{figure}[tbp]
    \begin{center}
	\includegraphics[scale=0.5]{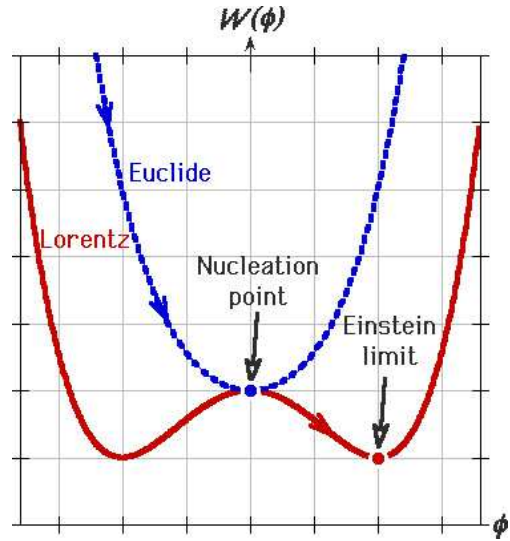}
    \end{center}
    \caption{Geodetic brane nucleation: From the Euclidean
    no-boundary initial conditions to the Einstein limit.}
    \label{Fig1}
\end{figure}

\medskip
Which potential actually governs the imaginary time
evolution of $\phi_{E}$?
Traditionally, we have been accustomed with the upside-down
potential $W_{E}(\phi_{E})=-W(\phi_{E})$, but this is
not the case here.
Euclidizing the time derivatives in the scalar field equation,
and simultaneously taking care of $\phi\rightarrow i\phi_{E}$,
brings us back to
\begin{equation}
    \phi_{E}^{\prime\prime}+
    3\frac{a_{E}\prime}{a_{E}}\phi_{E}^{\prime}+
    \frac{\partial W_{E}}{\partial\phi_{E}} = 0 ~,
    \label{Escalar}
\end{equation}
only with $W_{E}(\phi_{E})=+W(i\phi_{E})$.
The resulting potential
\begin{equation}
    \fbox{${\displaystyle W_{E}(\phi_{E}) = \Lambda +
    \frac{3\Lambda^{1/2}k^{2}}{16\omega}
    \left(\phi_{E}^{2} +\frac{4\omega\Lambda^{1/2}}
    {9k^{2}}\right)^{2}}$}
\end{equation}
although being quartic, is strikingly \textit{not} of the
Higgs type.
Furthermore, as depicted in Fig.(\ref{Fig1}), the absolute
minimum of $W_{E}(\phi_{E})$ is \textit{tangent} to the
local maximum of $W(\phi)$.
This is by no means coincidental. 
$\phi=\phi_{E}=0$ is the only point where the
Euclidean to Lorentzian transition ($\equiv$ brane
nucleation\cite{HLV}) can take place.

We now attempt to go one step beyond de-Sitter inflation.
To do so, we would like to commit ourselves to a certain
type of scalar potentials, but soon realize that so far
we have not really decoded the principles underlying the
tenable eq.(\ref{Higgs}).
The main question is this: \textit{Why must $W(\phi)$
exhibit a quartic behavior, and is such a quartic potential
a mandatory ingredient of geodetic brane cosmology?}
The answer to this question is rooted, quite unexpectedly,
within the Hartle-Hawking no-boundary ansatz\cite{HH}.
We prove that by exclusively predicting a finite non-vanishing
total energy density at the origin, the quartic structure
of the potential actually chooses the no-boundary initial
conditions.
The smoothness of the Euclidean manifold at the origin dictates
the specific $\tau\rightarrow 0$ behavior $a_{E}(\tau) \simeq
\sqrt{k}\tau$, but may in principle allow for
$\displaystyle{b_{E}(\tau) \simeq \frac{p\sqrt{k}}{\tau^{j-1}}}$.
Now, assuming the asymptotic power behavior 
\begin{equation}
    W(\phi) \simeq \lambda\phi^{N} \quad (\lambda>0) ~,
\end{equation}
the scalar equation of motion eq.(\ref{Escalar}) can
be fulfilled (to the leading order) only provided
\begin{equation}
    jN=2(j+1) ~, \quad 
    N\lambda p^{(N-2)}+j(j-2)=0 ~.
\end{equation}
This in turn implies $\rho \sim \tau^{-2(j+1)}$ but
$\rho_{total}\sim\tau^{4(j-1)}$.
Consequently, fully consistent with our expectations,
$N$ gets uniquely fixed by insisting on approaching a
\textit{finite non-vanishing} total energy density limit
as $a_{E}\rightarrow 0$.
This singles out $j=1\Rightarrow N=4$.
The generalized no-boundary initial conditions then read
\begin{equation}
    a_{E} \simeq \sqrt{k}\tau ~, \quad
    b_{E} \simeq \frac{k}{4\lambda} ~,
    \label{initial}
\end{equation}
accompanied by the finite total energy density
\begin{equation}
    \fbox{$\displaystyle{\rho_{total}\simeq
    \left(\frac{4\omega \lambda}{3k^{2}}\right)^{2}}$}
\end{equation}
While the no-boundary initial conditions are in fact
$\omega$-independent, it is $\omega$ which fixes the finite
total energy density.
It is interesting to note that had we carried out a similar
calculation for an $n$-brane (we skip the proof due to
length limitation), we would have encountered the famous
scale invariant behavior
\begin{equation}
   W(\phi) \sim \phi^{\frac{2(n+1)}{n-1}}  ~, 
\end{equation}
which happens to be quartic if $n=3$.
This indicates that, within the framework of geodetic brane cosmology,
there exists a linkage between (the apparently disconnected ideas of)
Hawking-Hartle no-boundary proposal and global conformal invariance,
pointing presumably towards geodetic dilaton cosmology.
\begin{figure}[tbp]
    \begin{center}
	\begin{tabular}{cc}
	    \includegraphics[scale=0.25]{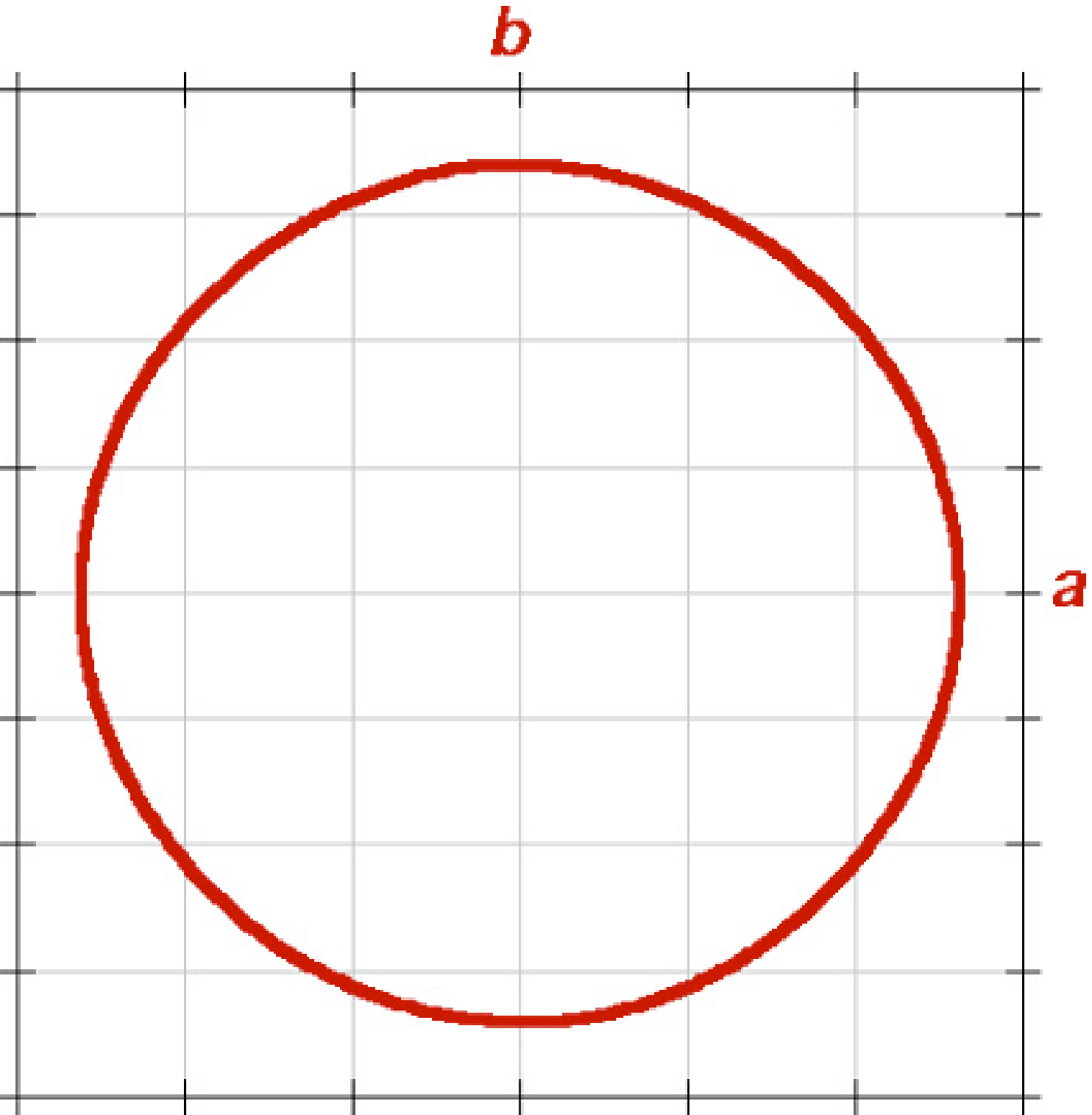} &
	    \includegraphics[scale=0.25]{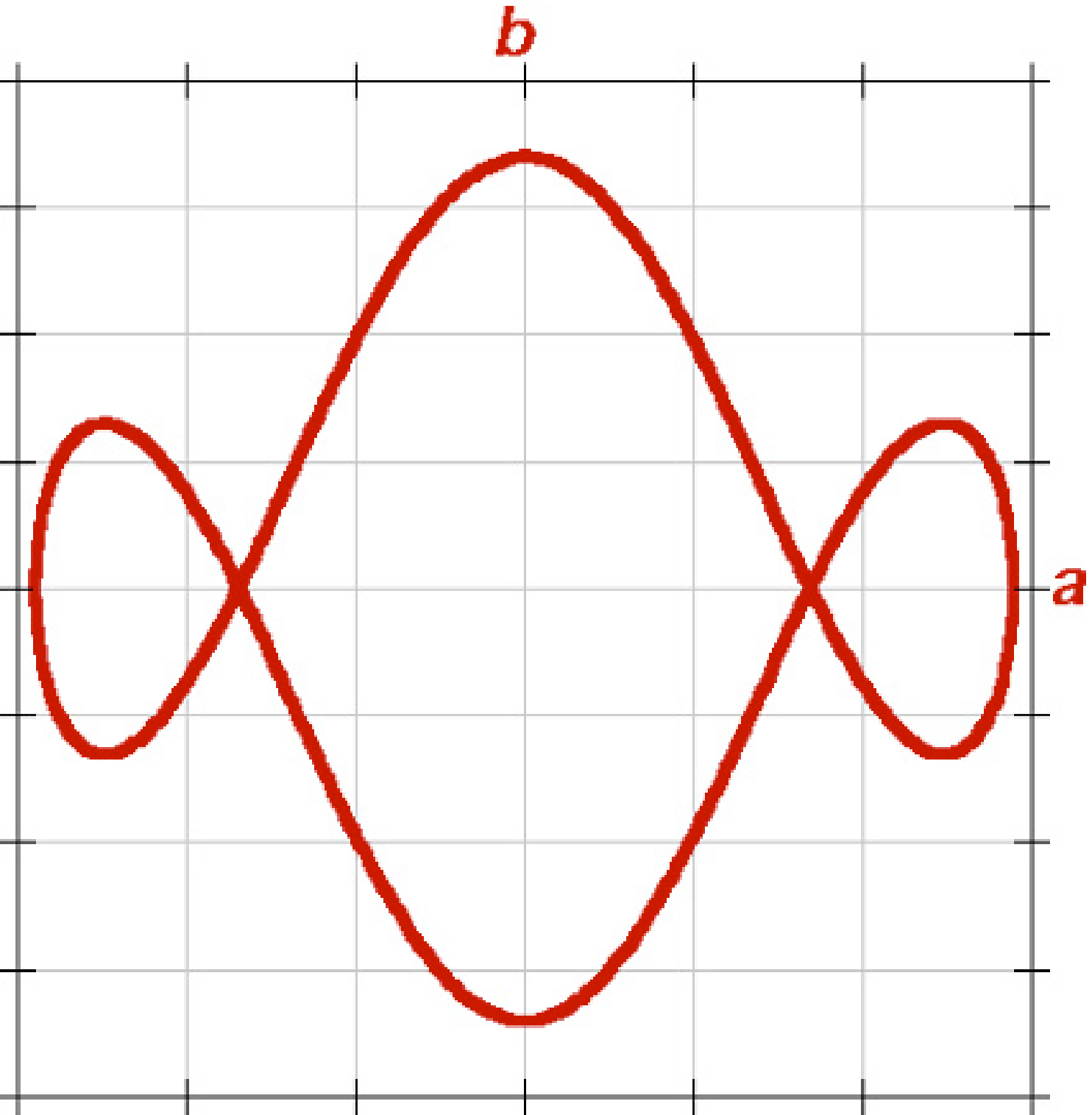}  \\
	    \includegraphics[scale=0.25]{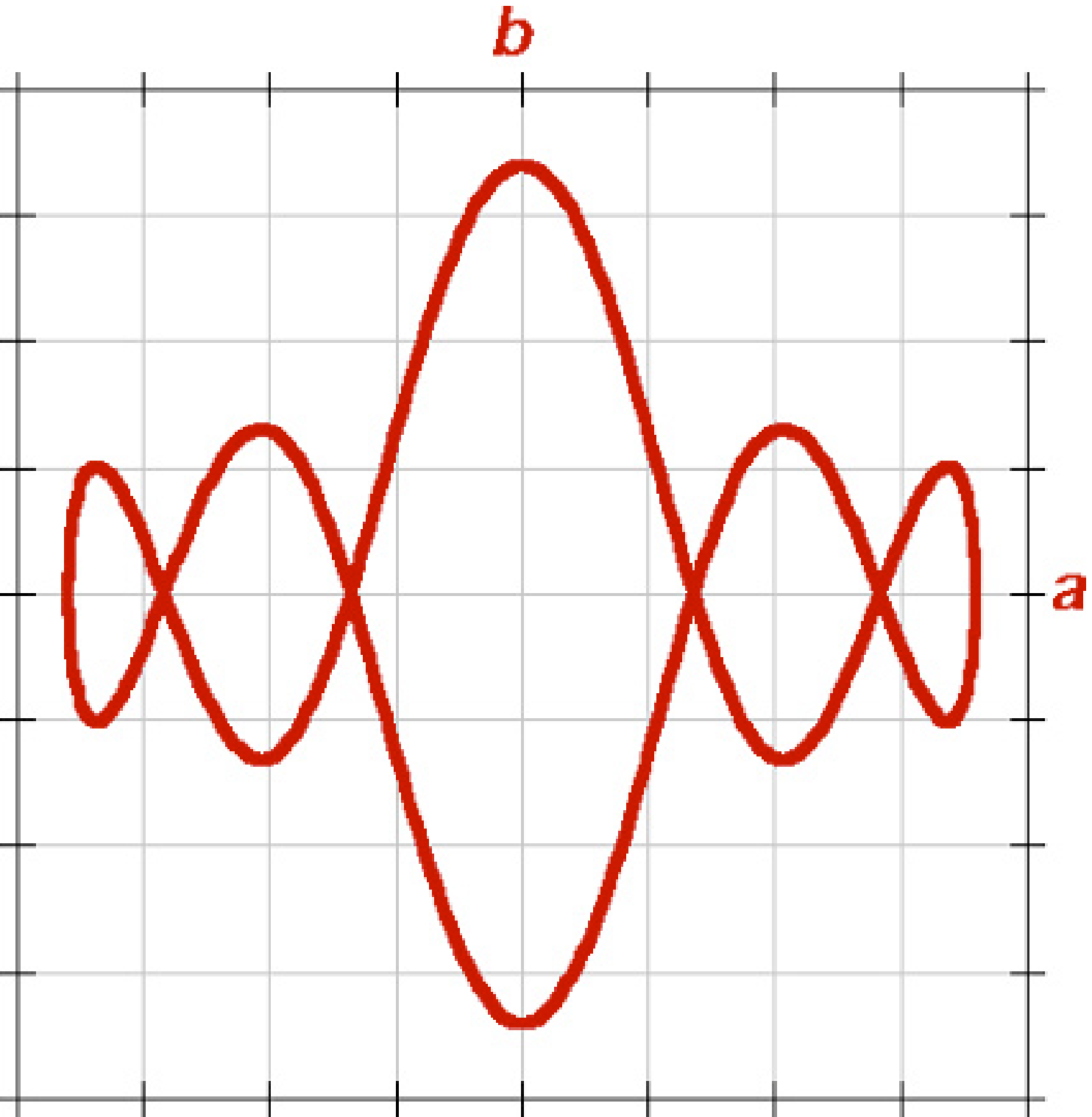} &
	    \includegraphics[scale=0.28]{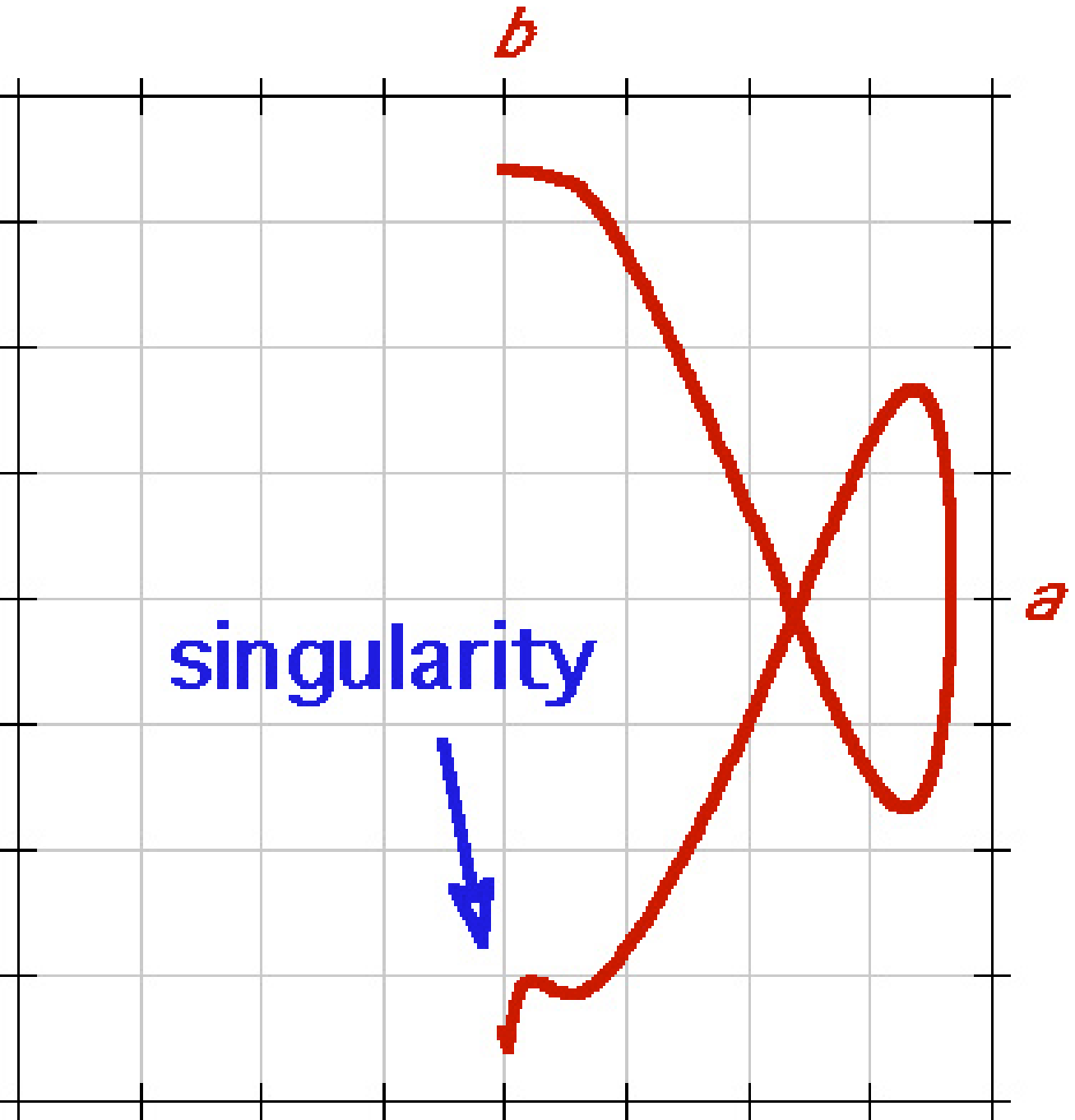}  \\
	\end{tabular}
    \end{center}
    \caption{Imaginary time periodicity and energy density
    regularity are demonstrated, for $\omega_{1,3,5}$, by means
    of closed trajectories in the $\{a_{E},b_{E}\}$ plane. If
    $\omega\neq\omega_{n}$, a $\cos(\ln\epsilon)$-type singularity
    is developed upon returning to the origin.}
    \label{Fig2}
\end{figure}
Finally, on realistic grounds, while adopting the quartic Higgs
potential, it makes sense to exercise the option of setting
$W_{min}(\phi)$ to zero.
The price for eliminating the residual cosmological constant from
the Einstein limit is a \textit{finite} (yet enhanced in comparison
with standard cosmology) amount of inflation.
Subject to the consistent no-boundary initial conditions
eq.(\ref{initial}), the classical Euclidean evolution is fully
determined once the conserved bulk energy $\omega$ gets
specified.
Naturally, this provokes a new set of questions:
(i) Does the global structure of the Euclidean manifold still exhibit
imaginary time periodicity?
(ii) Under what circumstances, if any, does the total energy density
evolve free of singularity?
(iii) Can our nucleation conditions $a_{E}^{\prime}=\phi_{E}=0$, or
else Coleman-Dellucia\cite{CD} conditions
$a_{E}^{\prime}=\phi_{E}^{\prime}=0$, be met at some finite
Euclidean time $\tau_{0}$?

\medskip
We claim, skipping the analytic proof (to be published elsewhere)
due to length limitation, that \textit{$\tau$-periodicity,
$a_{E},b_{E}$-regularity, and spontaneous nucleation, share the
one and the same origin.}
They can all be simultaneously achieved provided
$\omega=\omega_{n}$ is properly quantized, in agreement with
some previous WKB approximation\cite{AD}.
The integer $n$ counts the total number of times the proper
scalar field $b_{E}$ crosses the absolute minimum of the potential
during half a period.
For the $n$-odd case of interest ($n$-even is associated with
Coleman-Dellucia), reflecting the interplay of two periodicities, we
encounter (see Fig. 2) $n$-loop \textit{closed} trajectories in the
$\{a_{E},b_{E}\}$ plane which resemble the Lissajous figures.
Notice that the Euclidean de-Sitter configuration eq.(\ref{circle})
clearly belongs to the $n=1$ category.


\begin{references}
\bibitem{RS}
    L. Randall and R. Sudrum, Phys. Rev. Lett. 83, 3370 (1999).
\bibitem{RT}
    T. Regge and C. Teitelboim, in Proc. Marcel Grossman (Trieste),
      77 (1975).
\bibitem{E}
    E. Kasner, Am. Jour. Math. 43, 126, 130 (1921);
    C. Fronsdal, Phys. Rev. 116, 778 (1959)
    Y. Ne'eman and J. Rosen, Ann. of Phys. 31, 391 (1965);
    J. Rosen, Rev. Mod. Phys. 37,204 (1965);
    H.F. Goenner, in  \textit{General Relativity and Gravitation}
      (ed. A. Held, Plenum), 441 (1980);
    S. Deser and O. Levin, Phy. Rev. D59, 064004 (1999).
\bibitem{GW}
    G.W. Gibbons and D.L. Wiltshire, Nucl. Phys. B287, 717 (1987);
    R. Basu, A.H. Guth and A. Vilenkin, Phys. Rev. D44, 340 (1991).
\bibitem{RTgr}
    S. Deser, F.A.E. Pirani, and D.C. Robinson, Phys. Rev.
      D14, 3301 (1976).
    V. Tapia, Class. Quan. Grav. 6, L49 (1989);
    M. Pavsic, Phys. Lett. A107, 66 (1985);
    D. Maia, Class. Quan. Grav. 6, 173 (1989);
    I.A. Bandos, Mod. Phys. Lett. A12, 799 (1997).
\bibitem{AD}
    A. Davidson, Class. Quan. Grav. 16, 653 (1999);
    A. Davidson, D. Karasik, and Y. Lederer,
    Class. Quant. Grav. 16, 1349 (1999).   
\bibitem{CL}
    S. Coleman and K. Lee, Nucl. Phys. B329, 387 (1990). 
\bibitem{HH}
    J.B. Hartle and S.W. Hawking, Phys. Rev. D28,2960 (1983);
    J.J. Halliwell and S.W. Hawking, Phys. Rev. D31, 1777 (1985).
\bibitem{HLV}
    S.W. Hawking and I.G. Moss, Phys. Lett. 110B, 35 (1982);
    A.D. Linde, Sov. Phys. JETP 60, 211 (1984);
    A. Vilenkin, Phys. Lett. 117B, 25 (1982);
    A. Vilenkin, Phys. Rev. D30, 509 (1984);
    N. Turok and S.W. Hawking, Phys. Lett. B425, 25 (1998);
    A.D. Linde, Phys. Rev. D58, 083514 (1998) ;
    A. Vilenkin, Phys. Rev. D57, 7069 (1998).
\bibitem{CD}
    S. Coleman and F. De Luccia, Phys. Rev. D21, 3305 (1980).
\end{references}
\end{document}